\documentclass[aps,twocolumn,superscriptaddress,showpacs,showkeys]{revtex4}

\newcommand{\bastar}{\begin{eqnarray*}}
\newcommand{\eastar}{\end{eqnarray*}}
\newskip\humongous \humongous=0pt plus 1000pt minus 1000pt

\newif\ifdtup

\relax

\newcommand{\be}{\begin{equation}}
\newcommand{\ee}{\end{equation}}
\newcommand{\bea}{\begin{eqnarray}}
\newcommand{\eea}{\end{eqnarray}}
\newcommand{\X}{{\vec X}}
\newcommand{\pro}{\partial}
\newcommand{\n}{\hat n}
\newcommand{\oneg}{\displaystyle\frac{1}{g}}

\newcommand{\D}{{\hat D}}

\newcommand{\valpha}{{\vec \alpha}}

\newcommand{\dfrac}{\displaystyle\frac}
\newcommand{\ba}{\begin{array}}
\newcommand{\ea}{\end{array}}

\newcommand{\nn}{\nonumber}
\newcommand{\mn}{{\mu\nu}}

\newcommand{\Int}{\displaystyle\int}
\newcommand{\ab}{{\alpha\beta}}
\begin{document}

\title{Nucleon Momentum Decomposition in QCD}
\bigskip

\author{Y. M. Cho}
\email{ymcho@unist.ac.kr}
\affiliation{School of Electrical and Computer Engineering  \\
Ulsan National Institute of Science and Technology, Ulsan 689-798, Korea}
\affiliation{School of Physics and Astronomy,
Seoul National University, Seoul 151-747, Korea}
\author{Mo-Lin Ge}
\affiliation{Theoretical Physics Division, Chern Institute of Mathematics \\
Nankai University, Tianjin 300071, China}
\author{D. G. Pak}
\affiliation{Bogoliubov Laboratory of Theoretical Physics, Joint Institute for Nuclear Research \\
Dubna, Moscow region, 141980, Russia}
\author{Pengming Zhang}
\affiliation{Institute of Modern Physics,
Chinese Academy of Science, Lanzhou 730000, China}

\begin{abstract}
\noindent
\textbf{Abstract:}
~Based on the gauge invariant quark canonical momentum we construct two theoretically 
possible decompositions of nucleon momentum to those of quarks and gluons. We 
predict that either $6\%$ or $21\%$ of nucleon momentum is carried by gluons, 
depending on what type of gluons are in nucleons. We clarify the existing confusions 
on this problem and discuss the physical implications of our result on the proton 
spin crisis problem. 
\end{abstract}
\vspace{0.3cm}

\maketitle

An important problem in nuclear physics is to find out how much fraction of nucleon
momentum is carried by the gluons. It has generally been believed
that gluons carry about a half of nucleon momentum \cite{pol}. But recently
there has been a new assertion that only about one-fifth of the nucleon
momentum should be carried by the gluons \cite{chen}. This has created considerable
controversy and confusion in the literature \cite{ji,waka}.

To resolve this problem one has to know how to decompose the momentum
of nucleon to those of its constituents. At the first glance this problem 
seems to be simple enough. But in gauge theories it is very difficult 
to obtain a gauge invariant decomposition of momentum or spin to those of 
the constituents. In fact it has long been suggested that this is impossible 
in gauge theories. The reason is that the gauge interaction
makes a gauge invariant decomposition of the total momentum (and spin) to 
those of the constituents very difficult \cite{jauch,sak}.
{\it The purpose of this Letter is to clarify the confusion on this problem and
provide new nucleon momentum decompositions to predict the fraction of
gluon momentum in nucleons.}

To understand the problem, consider the canonical decomposition of
the momentum of positronium in QED
\bea
&P_\mu^{(qed)}= P_\mu^e +P_\mu^{\gamma}
= i\Int {\bar \psi} \gamma^0  \pro_\mu \psi d^3 x  \nn\\
&+\Int [(\pro_\mu A_\alpha)F^{\alpha 0} +\dfrac{1}{4} \delta^0_\mu F_{\ab}^2]d^3x.
\label{ap1}
\eea
This does provide a decomposition of momentum to those of the constituents, 
but is not gauge invariant. We can change it to the popular gauge invariant 
decomposition adding a surface term \cite{chen,waka}
\bea
&P_\mu^{(qed)}= {\bar P}_\mu^e +{\bar P}_\mu^{\gamma}
=i\Int {\bar \psi} \gamma^0  D_\mu \psi d^3 x \nn\\
&+\Int (F_{\mu\alpha} F^{\alpha 0}+\dfrac{1}{4}\delta_\mu^0 F_{\ab}^2) d^3x.
\label{ap2}
\eea
But this also may not be the desired decomposition because the first term 
involves both electron and photon. 

The problem stems from the fact that charged particles have two momentums, 
the ``canonical'' one given by $-i\pro_\mu$ and the ``kinematic'' one given 
by $-iD_\mu$, but neither is suitable for the momentum decomposition of 
composite particles \cite{jauch,sak}. This is because the canonical momentum is 
not gauge invariant, and the kinematic momentum contains the gauge field.
Moreover, there are actually two different issues in this problem. The first
is theoretical: How to make a gauge invariant decomposition of the total momentum. 
The second is experimental: How to make a measurable (and gauge invariant) 
decomposition of the total momentum. 
This is more subtle because here we must figure out what are the measurable 
momentums of the constituents. 

To obtain a gauge invariant decomposition of the positronium momentum, 
we first decompose the photon field to the vacuum part $\Omega_\mu$ and 
the physical part $X_\mu$ \cite{cho1},
\bea
&A_\mu= \Omega_\mu +X_\mu,
~~~\Omega_\mu=\pro_\mu \theta,~~~\pro_\mu X_\mu=0.
\label{ad}
\eea
Notice that this decomposition is gauge independent. Moreover, the gauge
transformation affects only the pure gauge part, so that the physical part
remains gauge invariant. In particular, the physical part
here becomes a Lorentz covariant four-vector, so that we can make
$X_0=0$ choosing a proper Lorentz frame (as far as $X_\mu$ is space-like).

Now, adding a surface term to (\ref{ap2}), we can easily change it to \cite{waka}
\bea
&P_\mu= i\Int {\bar \psi} \gamma^0  {\bar D}_\mu \psi d^3 x
+\Int[(\pro_\mu X_\alpha) F^{ \alpha 0} \nn\\
&+\dfrac{1}{4} \delta_\mu^0 F_{\ab}^2]d^3x,
~~~~~{\bar D}_\mu=\pro_\mu-ie\Omega_\mu.
\label{ap3}
\eea
Unlike (\ref{ap1}) or (\ref{ap2}), each term now is gauge invariant and
at the same time involves only one constituent. So, theoretically it does
describe a gauge invariant decomposition of the total momentum. 

It has generally been believed that the kinematic momentum is what 
experiments measure, because this is gauge invariant. This has made
(\ref{ap2}) a popular decomposition. But here we have shown 
that the canonical momentum can also be expressed by a covariant 
derivative. So there are actually two gauge invariant momentums 
that we can construct and thus can possibly measure. If so, which 
momentum is measurable and why is that so? 

Classically it appears that the conserved momentum of a charged particle 
moving in an electromagnetic field is the sum of the kinematic momentum of 
the particle and the electromagnetic momentum (the Poynting vector) \cite{sak}. 
This favors the kinematic momentum. But quantum mechanically the kinematic 
momentum operators do not satisfy the canonical momentum commutation relation, 
since they do not commute. Moreover, the canonical momentum $\bar D_\mu$ 
defined by the vacuum potential is gauge invariant 
and at the same time satisfies the canonical commutation relation. This 
strongly implies that (\ref{ap3}) is the correct momentum decomposition. 

In QCD the conserved momentum obtained by Noether's theorem is given by
\bea
&P_\mu^{(qcd)}= i\Int {\bar \psi} \gamma^0  \pro_\mu \psi d^3 x  \nn\\
&+\Int [(\pro_\mu \vec A_\alpha) \cdot \vec F^{\alpha 0}
+\dfrac{1}{4} \delta_\mu^0 \vec F_{\ab}^2]d^3x.
\label{np1}
\eea
Adding a surface term we can change it to the popular gauge invariant decomposition
\bea
&P_\mu^{(qcd)}= i\Int {\bar \psi} \gamma^0  D_\mu \psi d^3 x  \nn\\
&+\Int [\vec F_{\mu \alpha} \cdot \vec F^{\alpha 0}
+\dfrac{1}{4} \delta_\mu^0 \vec F_{\ab}^2]d^3x.
\label{np2}
\eea
But again the first term contains quarks and gluons. To cure this defect we first 
have to find out the gauge covariant canonical momentum operator which does not 
include gluons. 

To construct such momentum operator we must have a gauge independent decomposition 
of the non-Abelian gauge potential to the vacuum part $\hat \Omega_\mu$
and the physical part $\vec Z_\mu$ similar to (\ref{ad}). 
Consider SU(2) QCD for simplicity, and let $\n_i~(i=1,2,3)$ be a gauge covariant
right-handed orthonormal basis in SU(2) space. Then imposing the vacuum condition
to the potential
\bea
&\forall_i~~~D_\mu \n_i=(\pro_\mu+g\vec A_\mu \times) \n_i=0.~~~(\hat n_i^2=1)
\label{vp}
\eea
we obtain the most general vacuum \cite{plb07},
\bea
\vec A_\mu \rightarrow \hat \Omega_\mu
= \dfrac{1}{2g} \epsilon_{ijk} (\n_i \cdot \pro_\mu \n_j)~\n_k.
\label{vps}
\eea
Next, we make the decomposition
\bea
&\vec A_\mu=\hat \Omega_\mu + \vec Z_\mu,
\label{vdec}
\eea
and find that under the gauge transformation we have
\bea
&\delta \hat \Omega_\mu= \dfrac{1}{g} \bar D_\mu \vec \alpha,
~~~~~\delta \vec Z_\mu= -\vec \alpha \times \vec Z_\mu, \nn\\
&\bar D_\mu=\pro_\mu+g \hat \Omega_\mu \times.
\eea
where $\vec \alpha$ is the (infinitesimal) gauge parameter. Notice that
$\vec Z_\mu$ becomes a Lorentz covariant (as well as gauge covariant)
four-vector. Finally, we impose the transversality condition to $\vec Z_\mu$
to make it physical
\bea
&\bar D_\mu \vec Z_\mu=0.
\eea
Notice that this is not a gauge condition, because it applies to any gauge.
Obviously this is the generalization of (\ref{ad}) to QCD which provides
the desired decomposition.

Now, we can modify (\ref{np1}) to
\bea
&P_\mu^{(qcd)}=i\Int {\bar \psi} \gamma^0  {\bar D}_\mu \psi d^3 x  \nn \\
&+\Int [(\bar D_\mu \vec Z_\alpha)\cdot \vec F^{\alpha 0}  
+\dfrac{1}{4} \delta_\mu^0 \vec F_{\ab}^2 ]d^3x,
\label{np3}
\eea
adding a surface term 
\bea
-\Int (\pro_\alpha \hat \Omega_\mu \cdot \vec F^{\alpha 0}) d^3x.
\eea
Clearly this provides a gauge invariant decomposition of total momentum to those
of the quarks and gluons. But this may not be the desired decomposition that
we are looking for. The reason is that QCD has two types of gluons, so that 
we have to figure out which become the constituents of nucleons \cite{prd80,prl81,cho1}. 

To see this one has to understand that QCD allows the Abelian decomposition 
which separates the gluons to the colorless binding gluons and the colored 
valence gluons gauge independently. Because of this we have two types of QCD, 
the restricted QCD (RCD) made of the binding gluons and the standard QCD made of 
all gluons. Moreover, QCD can be viewed as RCD which has the valence 
gluons as the colored source \cite{prd80,prl81}. So the valence gluons 
(just like the quarks) become another colored source which has to be confined. 
This means that RCD plays the crucual role in confinement, which is known as 
the Abelian dominance in QCD \cite{thooft,prd02}. This interpretation has 
been confirmed numerically in a series of lattice QCD calculations \cite{kondo1,kondo2}. 

Most importantly, the quark model of hadrons tells that nucleons (in particular
the low-lying nucleons) are made of three quarks which are colored, not quarks 
and colored gluons \cite{pdg}. The colored gluons make up glueballs. This implies 
that valence gluons have no place in nucleons. If so, only quarks and binding 
gluons should contribute to the nucleon momentum. But so far this important
point has completely been ignored.

To exclude the contribution of the valence gluons in (\ref{np2}) we have to
separate the valence gluons from the binding gluons. This can be done by
the Abelian decomposition \cite{prd80,prl81}. Let $\n=\n_3$ be the unit
isotriplet which selects the color direction at each space-time point, and
make the Abelian projection imposing the condition,
\bea
D_\mu \n = (\pro_\mu + g {\vec A}_\mu \times) \n = 0.~~~(\n^2=1)
\label{ap}
\eea
This selects the restricted potential
\bea
&\hat A_\mu =A_\mu \n - \oneg \n \times\pro_\mu \n.
~~~(A_\mu = \n\cdot \vec A_\mu)
\label{rp}
\eea
With this we have the Abelian decomposition \cite{prd80,prl81},
\bea
& \vec{A}_\mu = \hat A_\mu + \X_\mu,~~~(\hat{n} \cdot \vec{X}_\mu=0)
\label{adec}
\eea
where $\vec X_\mu$ is the valence potential. Notice that $\hat A_\mu$
by itself forms a connection space, so that under the (infinitesimal)
gauge transformation we have \cite{prd80,prl81}
\bea
&\delta \hat A_\mu = \oneg \D_\mu \valpha,
~~~~~\delta \X_\mu = - \valpha \times \X_\mu,
\eea
where $\D_\mu=\pro_\mu+g \hat A_\mu \times$. 
What is important about this decomposition is that it is gauge independent.
Once $\n$ is chosen, the decomposition follows automatically, independent
of the choice of a gauge.

Since $\hat A_\mu$ still contains the pure gauge degrees, we need 
to decompose it to the vacuum and physical parts,
\bea
&\hat A_\mu= \hat \Omega_\mu+ \vec B_\mu,
~~~~~\bar D_\mu \vec B_\mu=0,   \nn\\
&\vec B_\mu= B_\mu \n,~~~~B_\mu= A_\mu- \dfrac{1}{g} \n_1 \cdot \pro_\mu \n_2.
\label{adec2}
\eea
Notice that $\vec B_\mu$ (just like $\vec X_\mu$) is gauge and Lorentz covariant.
This is because both $\hat A_\mu$ and $\hat \Omega_\mu$ form a connection space
which is closed under the gauge transformation.

Now, it is straightforward to obtain the desired decomposition of nucleon momentum.
All we have to do is to replace $\vec Z_\mu$ by $\vec B_\mu$ and $\vec F_\mn$ to 
$\hat F_\mn$ in (\ref{np2}),
\bea
&P_\mu^{(rcd)}= i\Int {\bar \psi} \gamma^0 {\bar D}_\mu \psi d^3 x  \nn \\
&+ \Int[({\bar D}_{\mu} \vec B_{\alpha}) \cdot \hat F^{ \alpha 0} 
+\dfrac{1}{4} \delta^0_\mu \hat F_{\ab}^2] d^3 x,
\label{np4}
\eea
where $\vec B_\mu$ is the transverse binding gluon. Notice that (\ref{np3}) is
physically very similar to the QED expression (\ref{ap3}). 

Clearly we can derive (\ref{np3}) from RCD. In fact RCD has the conserved momentum
\bea
&P_\mu^{(rcd)}=i \Int {\bar \psi} \gamma^0 \pro_{\mu} \psi d^3 x  \nn \\
&+\Int [(\pro_{\mu} \hat A_{\alpha})\cdot \hat F^{\alpha 0}  
+\dfrac{1}{4} \delta^0_\mu \hat F_{\ab}^2] d^3 x.
\label{nprcd}
\eea
From this we can obtain (\ref{np3}) adding a surface term. 

Now, we come back to the difficult question: What are the quark and gluon 
momentums in nucleon? In QED we have two gauge invariant electron 
momentums, the canonical $-i\bar D_\mu$ and the kinematic $-iD_\mu$.
But in QCD we have three. To see this notice that we can express (\ref{np3}) by
\bea
&P_\mu^{(rcd)}=i \Int {\bar \psi} \gamma^0 {\hat D}_\mu \psi d^3 x  \nn \\
&+ \Int[\hat F_{\mu \alpha} \cdot \hat F^{ \alpha 0} 
+\dfrac{1}{4} \delta_\mu^0 \hat F_{\ab}^2 )d^3 x,
\label{np5}
\eea
adding a surface term. Notice that the first term represents the quark kinematic 
momentum, but this contains only the binding gluons. This tells that there are 
two gauge invariant quark kinematic momentums, $-i\hat D_\mu$ and $-iD_\mu$, on 
top of the gauge invariant canonical momentum $-i\bar D_\mu$. So here we can not 
just say that it is the kinematic momentum that experiments measure.

The above analysis tells us the followings. First, in gauge theories there is 
a gauge invariant decomposition of total momentum (and spin) of a composite
particle to those of the constituents, one in QED and two in QCD. 
But these decompositions involve the canonical momentum 
which may or may not be measurable by experiment. Second, if the canonical 
momentum is not measurable, there is no gauge invariant decomposition of total 
momentum (and spin) to those of constituents in the strict sense. But we still 
have ``partial'' decompositions which involve the kinematic momentum, 
again one in QED and two in QCD. The reason why we have two competing 
decompositions in QCD is because we have two types of gluons. If nucleons 
contain only the binding gluons, (\ref{np3}) or (\ref{np5}) must be the correct 
one. But if they contain all gluons, we must have (\ref{np2}) or (\ref{np4}).  

To find which decomposition is correct, suppose only the kinematic momentum 
is mesurable. In this case we have in the asymptotic limit \cite{pol}
\bea
P_\mu^g=\dfrac{2n_g}{2n_g+3 n_f} P_\mu^{tot}.
\label{gm1}
\eea
Now, the difference between (\ref{np5}) and the popular (\ref{np2}) is that 
(\ref{np5}) includes only the binding gluons ($n_g=2$) but (\ref{np2}) includes all 
gluons ($n_g=8$). So (\ref{np2}) gives the well-known prediction (with $n_f=5$ 
as usual) that about $51\%$ of nucleon momentum is carried by gluons \cite{pol}. 
In contrast (\ref{np5}) tells that only about $21\%$ of nucleon
momentum must be carried by gluons. 

Now, suppose only the canonical momentum is measurable. In this case (\ref{gm1})
must change, and it has been proposed that (\ref{gm1}) be replaced by \cite{chen}
\bea
P_\mu^g=\dfrac{n_g}{n_g+6 n_f} P_\mu^{tot}.
\eea
This should be confirmed by an idependent calculation, but suppose this is true.  
Then (\ref{np3}) which assumes that nucleons contain all gluons predicts that 
gluons carry about $21\%$ of nucleon momentum \cite{chen}. Notice the strange 
coincidence between this prediction and that of (\ref{np5}) based on 
(\ref{gm1}). They have the same prediction, but totally different 
physics. If nucleons have only binding gluons, however, (\ref{np4}) tells 
that only about $6\%$ of momentum must be carried by gluons. Notice that the 
fraction of gluon momentum becomes less if nucleons has only binding gluons, 
for obvious reason.
 
Exactly the same argument applies to the nucleon spin crisis problem \cite{cho1,ji2,chen2}. 
Here again there are three (one canonical and two kinematic) quark orbital angular momentums. 
Moreover, assuming that only the canonical angular momentum is measurable, we have two nucleon 
spin decompositions depending on which gluons are in nucleons \cite{cho1}. And only one of 
them can describe the correct nucleon spin decomposition. 
 
Independent of the details the essence of our analysis can be summarized 
as follows. {\it First, there exist more than one logically acceptable gauge 
invariant quark and gluon momentums in QCD.} Indeed quarks have three and 
gluons have four such momentums, as we have shown in (\ref{np2}), (\ref{np3}), 
(\ref{np4}), and (\ref{np5}). This is because QCD potential allows the vacuum 
and Abelian decompositions (\ref{vdec}) and (\ref{adec}), so that quarks have 
one canonical and two kinematic gauge invariant momentums \cite{prd80,prl81}. 
Clearly this is against the common wisdom \cite{ji,chen,waka,ji2,chen2}. 

{\it Second, we must know which gluons are in nucleons to have a correct momentum 
(and spin) decomposition.}  So far this point has completely been ignored, because 
it has always been believed that all gluons are in nucleons \cite{ji,chen,waka,ji2,chen2}. 
But the Abelian decomposition tells that QCD has two types of gluons, and the quark 
model implies that only the binding gluons are in nucleons \cite{prd80,prl81,pdg}. 
Certainly this is a very interesting new idea which is totally different from 
the standard belief, and it is important to find out which gluons are in nucleons. 
In this Letter we showed that we can tell which view is correct by measuring 
the gluon momentum in nucleons. 

We hope that our analysis will help to settle the current controversies
on nucleon momentum and spin decomposition \cite{chen,ji,waka,ji2,chen2}. 
A detailed discussion on this and related issues will be presented 
elsewhere \cite{cho1,cho}.

{\bf ACKNOWLEDGEMENT}

The work is supported in part by National Research Foundation
(Grant 2010-002-1564) of Korea and by Natural Science Foundation
(Grants 10604024, 11035006, and 11075077) of China.

\end{document}